# Teaching computational thinking to space science students

*Robert Jeffrey*[1][2], *Megan Lundy*[2], *Deirdre Coffey*[2], *Sheila McBreen*[2], *Antonio Martin-Carrillo*[2], *Lorraine Hanlon*[2]

**Abstract**

Computational thinking is a key skill for space science graduates, who must apply advanced problem-solving skills to model complex systems, analyse big data sets, and develop control software for mission-critical space systems. We describe our work using Design Thinking to understand the challenges that students face in learning these skills. In the MSc Space Science & Technology at University College Dublin, we have used insights from this process to develop new teaching strategies, including improved assessment rubrics, supported by workshops promoting collaborative programming techniques. We argue that postgraduate-level space science courses play a valuable role in developing more advanced computational skills in early-career space scientists.

**Acronyms/Abbreviations**

*UCD*　　University College, Dublin

*SS&T*　　Space Science & Technology

## 1. Introduction

Computational thinking has been identified as a key skill for 21st century graduates. It refers to the ways we think when we design computer programs to solve problems [1] [2]. This should be distinguished from "coding" or "computing" [3], which means implementing a solution in a specific programming language.

While computational thinking is an increasingly influential idea in education [3] [4], it has always played a key role in solving problems in space science. Modern space scientists will use it for Earth observation, data analysis, and flight system control, with space software a major area of growth in the space industry [10]. However, little has been written on how space science education helps early-career space scientists to develop these skills.

### 1.1. Computation & the MSc Space Science & Technology at UCD

The MSc in Space Science & Technology (SS&T) at University College Dublin (UCD) is a taught program designed to prepare science and engineering graduates for careers in the global space sector.

A typical cohort consists of 12–16 students, most of whom are recent graduates from Irish universities. Typically, 20–30% of the class are female. Approximately one third of the class are international students and about 10% join after a period working in industry. Most students have degrees in physics or astrophysics (about 60%) or engineering (about 30%, usually aerospace or electrical engineering).

The 12-month course consists of a total of 90 ECTS credits. It includes classroom-based modules covering the space environment, applications of space science, and professional development, as well as optional modules on Earth-observation, climate physics, advanced astronomy and astrophysics, and data science. Three 10-credit laboratory or project-based modules cover space detectors, CubeSats, applied systems engineering and space mission design. A final 30-credit 12-week internship with a space agency, company, or research group leads to a minor thesis and presentation.

---

[1] Corresponding author: Robert.Jeffrey@ucd.ie
[2] School of Physics & UCD Centre for Space Research, University College Dublin, Ireland

Programming plays important and varied roles in many of these modules. Students write short programs to do calculations on homework assignments, and develop longer, complex programs to control complex space systems. They write data processing pipelines to calibrate and characterise gamma-ray detectors in the Space Detector Lab. They design and simulate space telescopes. In the Satellite Subsystems Laboratory, they write software to interface to our CubeSat simulator, EduCube [5], and to control their own "TupperSats" – Raspberry Pi-based experimental payloads that they fly on high-altitude balloons [6].

While computational ability is an essential skill in the SS&T course, it is often not an explicitly assessed learning outcome. For example, the learning outcomes of the TupperSat project focus on understanding the space mission life cycle and systems-engineering processes. Since space project teams need software expertise as much they need mathematical or written skills, students need appropriate support to develop these computational skills.

We encourage our students to use Python to solve these problems, and students need to use advanced programming techniques, including handling large data sets, concurrency, object orientation, and exception handling. To solve problems of this scope and complexity, students must also learn to think clearly and creatively about what they are doing: they must learn computational thinking, as well as how to code.

From talking to our alumni and employers, we know that our students value the computational skills that they develop during the course as they move into industry. But we also know that they find the learning curve steep, with expectations set far higher than they are used to as undergraduates. We see this as instructors: often, students' progress early in the course is slowed to learn these core skills.

This sets the aim of this work: to better understand our students' needs and challenges developing the level of programming and computational skill needed to succeed on the course and in the space sector. We can do this with a user-focused, design thinking framework.

## 2. Methods — Design Thinking

Design Thinking is a creative problem-solving approach used in industry and education to improve user experiences. It is often framed as a sequence of stages or mindsets: "empathise, define the problem, ideate, prototype, test" [7], or "inspiration, ideation, implementation" [8]. These all capture a general principle: you must understand your user before you can understand their problems, and you must understand the problem before you can solve it.

Using this idea, we divided our work into three steps: understanding our students, defining the problem, and implementing solutions.

## 3. Step I — Understanding Our Students

The first step in our design process is to empathise with our students, to understand their needs, views and experiences on the MSc SS&T. To do this, we surveyed students who completed the course between 2018 and 2021.

The anonymised questionnaire consisted of 33 questions divided into sections covering students' prior experience, the course itself, and their reflections looking back from their current career position. Some questions were posed as (numerical or verbal) rating scales, (eg., asking students to rate their confidence in a skill), but most used more open-ended written responses, to elicit students' experiences or perceptions.

The 4 cohorts contacted included 56 students, and we received responses from 29 students. We reviewed the responses with respect to several key questions:

1. what do incoming students know?
2. what do students do after the course?
3. what do students find helpful?
4. what do students find challenging?
5. what do students expect on the course?

We used students' quantitative responses (as shown, for example, in Figures 1–3), supported by select quotations from their written responses. We focused on identifying common themes and challenges from across the written responses by affinity mapping [9].

### 3.1. What do incoming students know?

All respondents reported some prior programming experience, across a range of languages, but few students claim to have been confident programmers before joining the course (Fig. 1). One respondent specifically noted that they had been "over-confident" in their abilities, while another "didn't realise that [they] knew as little as [they] did".

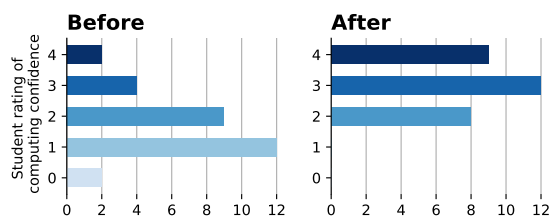

**Figure 1. Space-science students self-assessed programming confidence before and after the MSc SS&T (scale 0 - 4)**

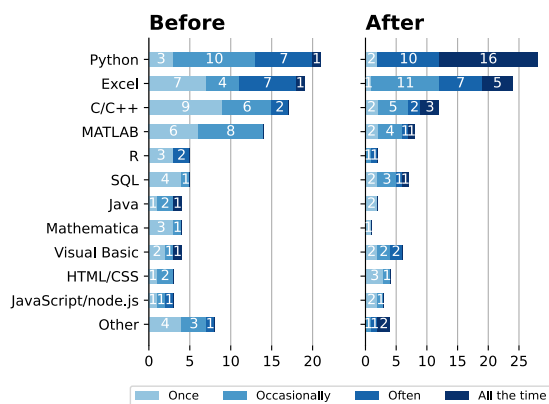

**Figure 2. Programming languages used by our incoming students, and graduates.**

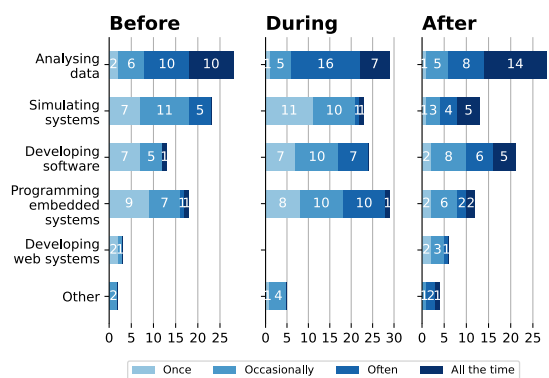

**Figure 3. What do space-science students use computing for?**

Nearly three quarters of students had used Python (Fig. 2), mostly for processing, analysis, and visualisation of laboratory data, or as part of an undergraduate research project. Almost two thirds had used C or C++ (often with Arduino microcontrollers), but students were less familiar with these languages. A small number had used technical or statistical software (eg., R, SAS).

Although most students (16 out of 29) had taken dedicated programming modules, this is not reflected in their written responses, which emphasise learning by writing code in labs. Formal programming classes appear disconnected from the rest of their learning; two respondents noted that after taking a course in C++ or Java, they "never used it again".

In general, most students' experience comes from data analysis or visualisation in labs. Figure 3 shows that this is the only programming application that students report as an often or always present part of their undergraduate experience. Most students have some experience with embedded programming, but usually only associated with a single project.

We supplemented this picture with a brief review of publicly available information on programming in physics and engineering at a selection of universities in Ireland and the UK. The general qualitative picture is that physics students' prior knowledge is narrow and deep, while engineering students' prior knowledge is broader and shallower. Students from a physics background typically have experience using Python for data analysis in undergraduate laboratories throughout their degree, with occasional courses in C/C++. Students from engineering courses tend to have used a wider range of languages (often including MATLAB, Excel or C), for a wider range of purposes (including modelling and numerical methods), but often only in the early years of their course.

### 3.2. What do students do after the course?

All respondents said that their confidence in their abilities increased after the course (Fig. 1). Figure 3 shows that computation is a routine part of their work. Three quarters of graduates use computers for data analysis often or all the time. Significant minorities of graduates use simulation or software development often or all the time, with a noticeable increase compared to undergraduate experiences (Fig. 3).

The course's emphasis on Python appears justified, as clearly the most popular language. It is used almost universally by respondents, with one noting that it was "considered a default requirement" when applying for jobs.

When asked to identify gaps in their learning, graduates want more experience with advanced technical skills. This includes a wider selection of languages (especially C++, although SQL, R and Ada were mentioned), advanced paradigms (especially object-orientation, which the course introduces briefly), and machine learning (which 5 students identified as a significant part of their career). Graduates reported that the course gave them a "much more realistic expectation of what to expect to do in the workplace", but they want to develop skills including a better understanding of the software development cycle, and of what professional, production code should look like.

*3.3. What do students find helpful?*

About half of responses identify the software development for their balloon experiment as the most useful part of the course, highlighting its scope and complexity ("the most complex coding project I had engaged with"), the need for robustness ("[creating] code that…would work every time"), the new technical skills learnt (particularly object-oriented programming and concurrency), and "the importance of clarity in code for collaborative programming".

Students mentioned "structuring code" most often as the most valuable skill that they developed during the course. They mention that this helps with "better layout", with writing "good code that… can be read easily", and with "trying to break…problems into smaller chunks". They identify its role in enabling collaboration, noting that "being able to explain code to others… is much easier when code is structured neatly". and that "compartmentalisation…simplifies comprehension for larger projects".

Generally, these comments about collaboration and structure suggest that graduates see a gap between the simple problems they meet as undergraduates and the more complex problems they face in industry or research, and that they need help to cross this gap.

*3.4. What do students find challenging?*

The responses show that students find the amount of new material and the steep learning curve challenging. A quarter of respondents identified "adapting to a relatively new language", knowing what level of ability was expected, or finding appropriate resources as a source of difficulty at the start of the course. A similar number of students identified difficulty learning more advanced skills (eg., model fitting, objects, and embedded systems).

Students also identified challenges in the step up to more complex and open problems, in which you "really had to think for yourself" to come up with solutions, and where the program structure needs to be considered as part of this.

*3.5. What do students expect on the course?*

Students generally appear to be surprised by the level and nature of the programming that they encounter on the course. 15 respondents mention that there was more than they expected, while only 1 said there was less (specifically, less low-level programming). Five respondents commented that they "ultimately really appreciated" the amount of programming on the course, suggesting that although they find the process (unexpectedly) difficult, they can see the benefit on reflection.

Four respondents said that the type of programming that they were asked to do was unexpected. They "expected to spend most… time on data analysis", but that the course "moved away from analysis". The "software development was a lot harder than [they] expected", but they feel that they can apply skills to "more real-world tasks now". This again suggests that undergraduate courses cover a narrower range of applications and skills than graduates use in the space sector.

## 4. Step II — Defining the Problem

In the next step in the design thinking process, we identify the problem to be solved. By reviewing the student responses, we have built a clearer picture of a typical space-science student's experience, wants and needs. Through our affinity mapping exercise, we then identified a set of emergent themes, each framed as a problem experienced by students:

1. what is good code? Students are unclear what makes code "good", and how to implement good practice.
2. managing expectations – students are surprised by the level and nature of programming required by the course.
3. learning the basics – students find it difficult to learn the Python language at the same time as course material.
4. finding support – students want help to find additional learning resources.
5. managing larger problems – students struggle to manage the amount of data generated in the Space Detector Lab, and to adapt to the larger project scope in the Satellite Subsystems lab.
6. learning advanced skills – students find it difficult to learn the more advanced programming techniques that they use on the course.

Addressing these problem statements will form the basis of our course development work.

## 5. Step III — Implementing Solutions

The last three stages in the design process — ideation, prototyping, and testing — cover finding and implementing solutions to the problem(s) that we have defined. Using these problem statements and drawing inspiration from the students' questionnaire responses, we have trialled several interventions within the course in the current 2021-22 academic year.

### 5.1. Improved Assessment Rubrics

We have introduced a new assessment rubric for code submissions, to address students' uncertainty about what makes good code.

The assessment rubric is based around 3 criteria: functionality, structure, and style. Each criterion assesses a distinct aspect of thinking about code. Functionality assesses how well the code does what it needs to do. The structure rubric assesses the organisation of code, with credit for code that is logical, flexible, and reusable, that uses compartmentalisation and abstraction appropriately, and that separates what is being done from how it is done. Style assesses how professionally the code is written, including readability, effective documentation, consistent styling, and writing idiomatic code.

By giving equal credit to these three areas, we encourage students to think about both what their code does and how it is put together.

### 5.2. "Writing Programs in Python" Workshop

We ran a 3-hour introductory workshop to support this rubric, illustrating our expectations using a series of paired-development exercises.

The workshop consists of 3 activities. First, students are placed in pairs to peer-review another student's solution to a short pre-class coding exercise. The instructor then leads a class discussion on helpful and unhelpful practices in programming. They introduce the assessment rubric, with a live demonstration to illustrate how to transform bad code to good. Finally, the class divides into pairs for a pair-programming exercise based on the popular "FizzBuzz" problem [11]. In this activity, pairs of students act alternately as programmer and reviewer in short (6 minute) programming sprints, with the reviewer guided to look at the structure and style of students' solutions.

Using collaborative exercises helps students learn to write code that communicates their intentions. Students see what makes good code by watching someone else write code, learning from the strengths and weaknesses of their practice, using the rubric as a guide. Introducing pair programming and code review also helps students to learn the professional skills needed to work as part of a software team, a skill which our graduates valued in their own careers.

## 6. Discussion

### 6.1. Who are our students?

We can use our empathetic research in Section 3 to build up a profile of a "typical" member of the Space Science & Technology cohort.

Our typical student has previously used Python (or possibly MATLAB) for data analysis but has very little formal computing education. They like that the course teaches them code structure and collaboration, and they like learning advanced topics (including object-oriented programming and machine learning). They find that they struggle with the learning curve at the start of the course, and they are unclear about what is expected of them as a programmer.

Of course, this profile comes with the obvious caveat that it does not attempt to capture the academic and social diversity that our students bring to the course, and we must be mindful that any solution based on it cannot be a one-size-fits-all answer. Nevertheless, it suggests that postgraduate space science courses should not be afraid to emphasise advanced computational skills, but cannot assume students will have more than basic familiarity with coding.

### 6.2. What special programming skills do space science students need?

There is a gap between the programming skills our students learn on undergraduate courses, and the skills they need in industry. We can see this in the number of responses identifying "structure" as the most important skill they learn from our course. This tells us that students' previous experiences may have given them the basic literacy needed to complete small data analysis tasks, but have not prepared them to think about and solve the larger and more varied software development and data analysis problems they meet in the space sector.

Postgraduate space-science courses have an important role here. As well as teaching students space-sector specific knowledge and skills, they introduce students to the more complex computational problems that they may encounter as graduates working in software teams. Indeed, when students talk about learning to structure code, they often mean learning how to think about code – that is, computational thinking.

*6.3. How well did it work? What next?*

The design thinking process encourages reflection and iteration, and there are lessons to be learnt from this exercise for future years.

The rubric was used throughout the first semester lab modules to guide students and give feedback. This has enabled more focused discussions with students as they developed their code, and simplified giving feedback.

From our (qualitative) observation of this year's cohort so far, students have taken onboard our emphasis on professionalism in their code development, suggesting that the emphasis on style and structure has worked. For example, we have seen more instances of students discussing their code together or using whiteboards and flow diagrams to plan out and structure their code before they write it.

Although students have a clearer understanding of the level that they will be expected to achieve, we have not yet addressed the challenges faced by those students learning to code with little or no prior experience. We expect students to prepare for the course by familiarising themselves with the fundamentals of the Python language, but find that this is done inconsistently. This is a harder problem to solve: the obvious solutions involve finding additional resources (by providing a pre-course training camp on Python), or compromising other parts of the course (by reducing space-science specific learning outcomes). We are looking at technology-enabled solutions to help incoming students reach a clearer common baseline.

Lastly, we have focused here on the needs and experiences of students and recent graduates. However, other stakeholders will need to contribute to developing best practice. Most notably, we will need industry input to identify the most useful technical and professional computational skills for new space-scientists.

## 7. Conclusions

Our work on teaching programming on the MSc Space Science & Technology at UCD provides a case study in using design thinking processes in education. This has helped us to identify some challenges our students face developing computational thinking skills as they move from higher education into industry, especially in understanding the higher standards, greater complexity, and wider variety of programming problems that they encounter as early-career space scientists. We have briefly described possible ways to use clearer expectations to smooth this transition, but this is an evolving area where best-practice has yet to emerge.

**Acknowledgements**

We thank UCD School of Physics for supporting this work, and the MSc SS&T students for their insights, engagement and enthusiasm.